\documentclass[10pt,twocolumn]{article}

\usepackage[margin=0.85in]{geometry}
\usepackage{newtxtext}
\usepackage{newtxmath}
\usepackage{microtype}

\usepackage{graphicx}
\usepackage{booktabs}
\usepackage{siunitx}
\usepackage{xcolor}
\usepackage{float}
\usepackage[hidelinks]{hyperref}
\hypersetup{
    colorlinks=true,      
    linkcolor=blue,       
    citecolor=blue,       
    urlcolor=blue         
}
\usepackage[numbers,sort&compress]{natbib}
\usepackage[font=small,labelfont=bf,labelsep=period]{caption}
\usepackage{subcaption}
\usepackage{tikz}
\usepackage{authblk}
\usepackage{titlesec}
\usepackage{cuted}

\newcounter{BalanceAtReference}
\titleformat{\section}
  {\large\bfseries}{\thesection}{0.75em}{}
\titleformat{\subsection}
  {\normalsize\bfseries}{\thesubsection}{0.75em}{}
\titleformat{\subsubsection}
  {\normalsize\itshape}{\thesubsubsection}{0.75em}{}

\title{Fish Navigate a Hydrodynamic Maze via \\ Yaw-Mediated Lateral Migration}

\author[1]{Michael A. Calicchia}
\author[1]{Rui Ni}
\affil[1]{Department of Mechanical Engineering, Johns Hopkins University}
\date{}

\newcommand{\legendcircleR}[3][red]{%
  \tikz[baseline=-0.6ex]{%
    \draw[black, fill=#1] (#2/2,0) circle (#3/2);
  }%
}

\newcommand{\legendcircleG}[3][green]{%
  \tikz[baseline=-0.6ex]{%
    \draw[black, fill=#1] (#2/2,0) circle (#3/2);
  }%
}

\makeatletter
\newcommand{\frontmatteritem}[2]{%
  \noindent\textbf{#1:} #2\par\vspace{0.8em}
}

\newcommand{\authorcontributions}[1]{\def\@authorcontributions{#1}}
\newcommand{\competingintereststatement}[1]{\def\@competingintereststatement{#1}}
\newcommand{\correspondingauthor}[1]{\def\@correspondingauthor{#1}}

\newcommand{\showauthorcontributions}{%
  \frontmatteritem{Author contributions}{\@authorcontributions}
}

\newcommand{\showcompetingintereststatement}{%
  \frontmatteritem{Competing interest statement}{\@competingintereststatement}
}

\newcommand{\showcorrespondingauthor}{%
  \frontmatteritem{Corresponding author}{\@correspondingauthor}
}
\makeatother

\authorcontributions{M.A.C and R.N designed the experiments and methodology. M.A.C. performed experiments, analyzed the data, made the visualizations, and wrote the original draft. M.A.C. and R.N. reviewed and edited the manuscript. R.N. was responsible for funding acquisition}

\competingintereststatement{The authors declare no competing interest.}

\correspondingauthor{Rui Ni, email: rui.ni@jhu.edu}

\begin{document}

\twocolumn[
\maketitle
\vspace{-1.0em}

{\small \textit{This is the author's peer-reviewed, accepted manuscript (postprint). The definitive version of record was published in the Proc. Natl. Acad. Sci. U.S.A. on Sept. 3, 2026, Vol. 123, No. 36, e2613565123, and is available online at: \url{https://www.pnas.org/doi/10.1073/pnas.2613565123}}}
\vspace{2.0em}

\frontmatteritem{Significance}{Fish often swim through fast, chaotic currents that are difficult and costly to resist. Instead of battling these flows head-on, we found that fish use a simple strategy: they slightly turn their bodies relative to the flow. This subtle change in orientation generates a passive lateral force that helps them reposition with little extra effort. By adopting these angled body orientations, fish also become more sensitive to subtle local flow differences, potentially helping them identify hydrodynamic refuges within otherwise challenging environments. Our findings reveal a previously unrecognized mechanism by which aquatic animals exploit, rather than overcome, environmental forces. Beyond advancing our understanding of animal movement, these insights may inspire strategies for energy-efficient navigation in autonomous vehicles and other engineered systems.}

\frontmatteritem{Abstract}{In aquatic environments, gradients in flow velocity and turbulence define a constantly shifting landscape that sets the physical constraints on propulsion and energy expenditure in fish. Successfully navigating these complex flows depends on the ability to sense and respond to subtle hydrodynamic cues in an energy-efficient manner. Revealing these mechanisms is central to understanding how fish identify and exploit favorable flow conditions. To this end, we designed a hydrodynamic maze with spatially varying mean flow and turbulence, creating a controlled heterogeneous flow landscape for understanding fish movement strategies. We found that fish escape energetically unfavorable regions by yawing their bodies relative to the flow. These yawed orientations appear to facilitate lateral migration through lift forces and may improve sensitivity to local flow variations. Notably, fish maintain body orientations near, but below, predicted stall conditions to exploit lift without a significant increase in drag. These findings provide insight into how fish navigate heterogeneous hydrodynamic environments, with potential relevance for both natural and engineered flow systems.}

\showauthorcontributions
\showcompetingintereststatement
\showcorrespondingauthor

\vspace{0.3em}
]

\section*{Introduction}

Navigating complex environments requires animals to make movement decisions based on cues that can be ambiguous, subtle, or even hidden. While much work has focused on how terrestrial animals, such as rodents \cite{Rosenberg2021, Saleem2023, Woronowicz2025}, ants \cite{Bisch-Knaden2001, McCreery2016}, and cockroaches \cite{Cowan2006, Baba2010, Wang2022}, avoid obstacles, a distinct but equally important question is how aquatic animals make navigation decisions when the relevant challenges arise from the surrounding flow, where velocity and turbulence are spatially and temporally varying, yet not directly visible or tangible.

Fish are one example of aquatic animals that have to cope with such invisible challenges. To do so, they rely on their lateral line to sense velocity and pressure gradients \cite{Montgomery2000, Ristroph2015, Dabiri2017, Oteiza2017}, which allows them to detect and avoid regions of high velocity \cite{Webb2006, Johansen2007, Liao2007, Webb2010, DiSanto2025} or elevated turbulence \citep{Smith2005,Smith2006,Cotel2006, Goettel2015, Smith2014, Hockley2014, Li2022_Animals, Liu2023}, both of which are associated with higher energetic costs \citep{Hinch1998, Pavlov2000, Enders2003, Roche2014, Maia2015, Agbeti2024, Zhang2024}.

\begin{figure*}[t!]
\centering
\includegraphics[width=\textwidth]{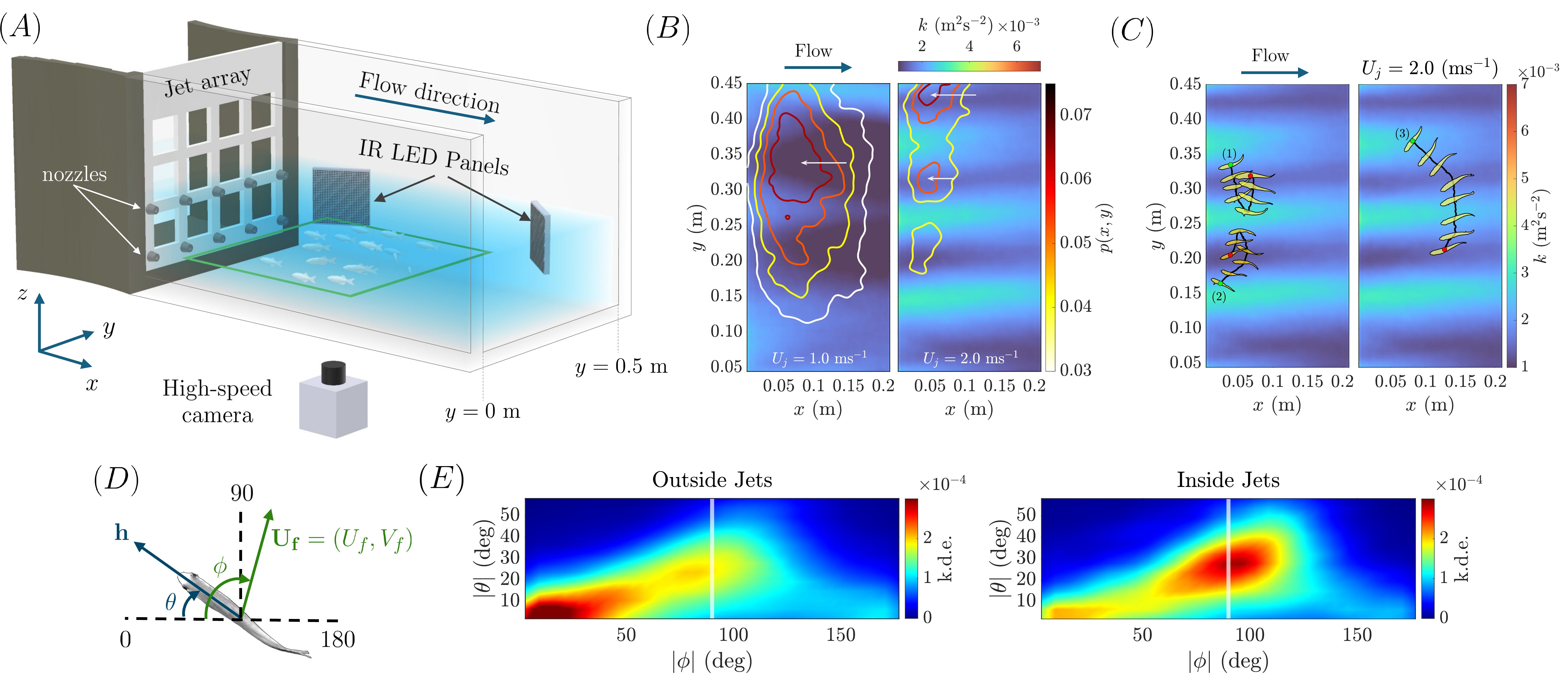}
\caption{\textbf{Fish navigate away from unfavorable flow conditions through lateral migration}. (A) Section of the experimental facility illustrating key features of the set-up (B) Spatial occupancy heatmaps $p(x,y)$ overlaid onto the turbulent kinetic energy profiles ($k$) showing preferential use of favorable flow regions and shifts in occupancy with changes in the hydrodynamic landscape. (C) Trajectories demonstrating fish using lateral migration to relocate to more favorable locations between jets ( \legendcircleG{1.2}{0.1} represents the starting point and \legendcircleR{1.2}{0.1} the end point). (D) Definition of orientation angle $\theta$ and direction of motion $\phi$. $\theta$ is measured clockwise from the streamwise direction to the heading direction vector ($\mathbf{h}$) and $\phi$ clockwise from the flow direction to the absolute fish velocity vector ($\mathbf{U_f} = (U_f, V_f)$) (E) K.d.e. approximation of the joint p.d.f. of $\theta$ and $\phi$ for fish located outside jets (locally lower mean velocity and turbulence) and inside jets (locally higher mean velocity and turbulence). The two white vertical lines denote $90^\circ$ indicating lateral motion.}
\label{fig:Fig1}
\end{figure*}

Understanding these navigation strategies is motivated by both fundamental research in animal locomotion and practical applications, such as improving the efficiency of bio-inspired marine robotics \cite{Jiang2019, Gunnarson2021, Othman2023, Koiri2025, Gunnarson2024, Hang2025, Jiao2025} or fishways \cite{Knapp2019, Silva2020, Silva2011, Silva2012, Silva2012b, Duguay2018, Silva2020}. Many researchers have developed behavioral models \cite{Mawer2019} to determine how flow stimuli, such as fluid acceleration \cite{Goodwin2014}, velocity \cite{Liao2024, Tan2019}, and turbulent kinetic energy \cite{Gao2016,Tan2019} influence fish movement decisions. Goodwin et al. \cite{Goodwin2014} showed that these models can reproduce passage‑efficiency patterns observed at different dams, whereas other studies \cite{Lemasson2008, Gao2016, Tan2019, Liao2024} have utilized similar models to replicate trajectories observed in experimental fishways. Together, these results indicate that the models capture essential real‑world structure in fish navigation and can be utilized to predict how changes to fishway designs will impact fish passage efficiency \cite{Zielinski2015, Liao2024}. 

Although most models implemented mechanisms to move fish away from high velocity or turbulence, they do not, however, explicitly resolve the exact sensing and actuation strategies adopted by fish. In particular, fish were often modeled as point particles or rigid bodies, thereby omitting body curvature and orientation dynamics, which are key factors in propulsion, maneuvering, and flow sensing.

To address this gap, we study the behavior of giant danio (\emph{Devario aequipinnatus}) as they navigate an experimental hydrodynamic maze that imposes controlled heterogeneity via a jet array. The jet array was designed to generate a repeating pattern of high-velocity, high-turbulence regions (energetically unfavorable) interspersed with lower-velocity, lower-turbulence regions (favorable). The alternating jets disrupt global flow cues, and because fish rely on local hydrodynamic information that does not predict conditions beyond nearby regions, navigation becomes inherently maze-like. In this context, we aim to uncover the mechanisms that enable fish to detect and efficiently reach the more energetically favorable regions of the flow.

\section*{Results and Discussion}

To examine how fish sense and navigate out of unfavorable flow conditions characterized by elevated velocity and turbulence, we constructed an experimental hydrodynamic maze with controlled spatial heterogeneity in flow using an array of jets. Details of the experimental facility and generated flow fields can be found in \cite{Calicchia2026} and in the supplementary material. A section of the facility highlighting the swimming volume and key components is shown in Fig. \ref{fig:Fig1}(A). The turbulent kinetic energy profile for two flow conditions are also shown in Fig. \ref{fig:Fig1}(B). The flow pattern at the intermediate jet speed shows alternating unfavorable regions (cyan) and favorable regions (blue), each spanning roughly three to five body widths. At this scale, fish primarily experience local flow conditions and cannot reliably infer the structure beyond their immediate surroundings.

\begin{figure*}[t!]
\centering
\includegraphics[width=\textwidth]{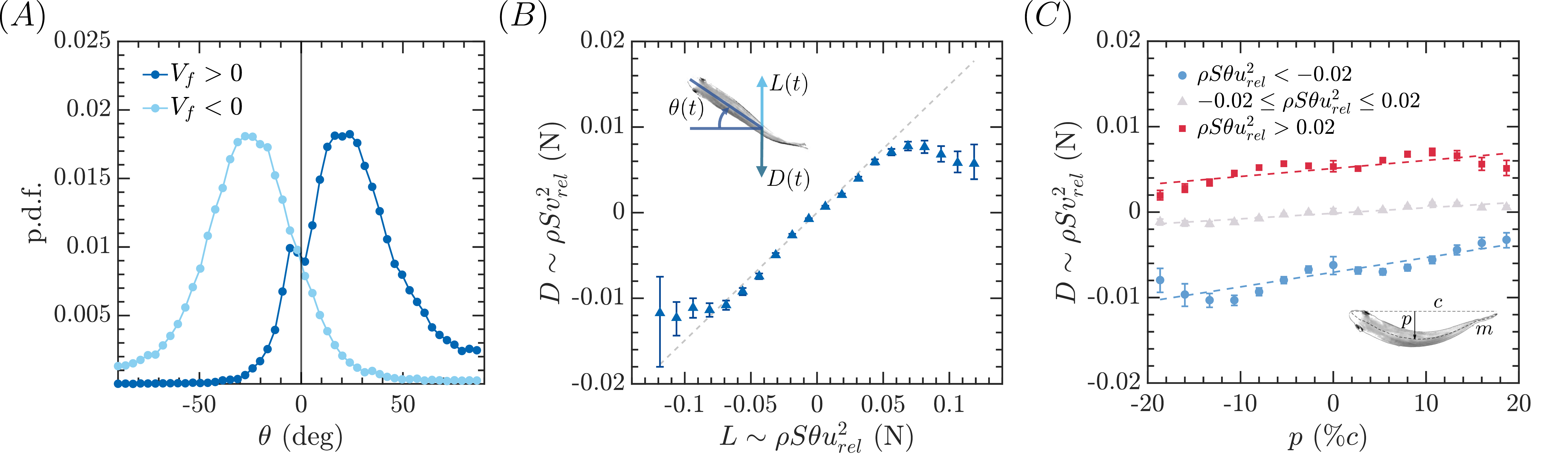}
\caption{\textbf{Lift-based scaling associated with lateral migration}. (A) P.d.f of the orientation angle conditioned on the lateral velocity, showing directional coupling between body orientation and lateral motion. (B) Lateral drag force versus lift force, showing broad agreement with the expected linear scaling (dashed line) supporting the proposed lift-based transport. (C) Lateral drag force versus maximum percent camber $p$, suggesting that body curvature influences the lateral lift force. A representative fish is also included showing the midline $m$, chord $c$ and maximum camber. Symbols in (B) and (C) denote the event-level mean; error bars indicate 95\% confidence intervals estimated using a hierarchical bootstrap over tracks and events. Track and event counts for each bin is reported in Table S2.}\label{fig:Fig2}
\end{figure*}

To quantify how fish distribute themselves within these heterogeneous flow fields, two-dimensional spatial occupancy maps $p(x,y)$ were computed from tracked fish positions to estimate the probability of observing a fish at each location and were overlaid onto the turbulent kinetic energy fields in Fig. \ref{fig:Fig1}(B). The occupancy maps indicate a modest preference for the upper boundary of the swimming area over the lower boundary. This asymmetry may reflect environmental differences between the two sides of the experimental set-up: the lower boundary was covered by a dark surface, whereas the upper boundary contained infrared light panels that may have been slightly more attractive to the fish. Importantly, the reconstructed velocity field does not extend all the way to the physical walls. Thus, the occupancy peak near the upper edge of the plotted flow field should not be interpreted as direct wall contact or thigmotaxis \cite{Colwill2011, Sharma2009}, as this location remains approximately 5-8 cm from the physical wall.

Furthermore, apart from this modest boundary-associated asymmetry, the occupancy patterns vary systematically with the flow conditions. At lower jet speeds, larger portions of the flow field remain within relatively favorable hydrodynamic conditions, allowing fish to distribute themselves over a broader region. As jet speed increases, the regions characterized by low velocity and low turbulence become smaller and more localized. Fish occupancy changes correspondingly, with the highest-occupancy regions becoming more spatially concentrated in the inter-jet regions. This tightening of the occupancy hotspot, indicated by the arrows, suggests that fish can detect and preferentially migrate toward energetically favorable regions of the flow.

Fig. \ref{fig:Fig1}(C) illustrates representative trajectories observed in the experiments. The first two trajectories show fish entering regions of high velocity and turbulence before reorienting and returning to the low-disturbance refuge between adjacent jets. The third trajectory demonstrates that large yaw angles relative to the incoming flow can generate sufficient lateral displacement for a fish to overshoot a jet and relocate to the next sheltered region.

These observations suggest that fish can effectively locate and remain within hydrodynamically favorable regions in complex flows, and that their ability to depart from unfavorable locations may be closely linked to body orientation. To test this, we measured the body orientation $\theta$ and resulting direction of motion $\phi$, as shown in Fig. \ref{fig:Fig1}(D). These angles are measured clockwise from the streamwise direction, such that $\theta = 0$, $\phi = 0$ corresponds to a fish aligned with and swimming directly against the flow. Fig. \ref{fig:Fig1}(E) shows the joint probability density function (p.d.f.) of $\theta$ and $\phi$ aggregated over all flow regimes when fish are in two different regions of the flow. A velocity threshold at fish centroid locations is used to determine when fish are inside and outside the jets. 

Both p.d.f.s appear to show a bimodal distribution with a primary (red) and secondary (orange) peak indicating that fish switch between two behavioral states: maintaining position or changing location. Outside the jets, the primary peak is located at $\theta \approx 0^\circ, \phi \approx 4^\circ$, indicating that fish typically align their bodies at small angles to the incoming flow and predominantly swim upstream, consistent with station-holding in preferred flow regions. Conversely, within the jets, this behavior is associated with the secondary peak, and a different dominant behavior is observed. The peak occurs at $\theta \approx 30^\circ, \phi \approx 97^\circ$, which reveals that fish preferentially orient at a finite body angle, but surprisingly do not rapidly swim out of the jets by moving in their heading direction. Instead, the peak near $ 90^\circ$ indicates that fish predominantly exit these regions by moving laterally, effectively sliding sideways relative to the flow, as demonstrated in the representative trajectories. 

These results suggest that the hydrodynamic conditions modulate the fish’s dominant behavioral state. Station-holding becomes the predominant state in more energetically favorable flow conditions, whereas lateral migration dominates in regions of the flow that are more energetically costly. 


The observed increase in orientation angle, together with the lateral motion, suggests that fish may exploit the lift force acting on the body to move laterally and escape the unfavorable hydrodynamic conditions. If lift contributes to fish migration, body orientation and lateral motion should be directionally coupled, and the resulting force should scale with body orientation and the square of relative velocity. To test this, we first examine whether yawing the body in a given direction is statistically associated with lateral movement in that same direction. 


Fig. \ref{fig:Fig2}(A) shows the p.d.f. of fish orientation conditioned on downstream motion when $V_f < 0$ or $V_f > 0$. In both cases, the distributions exhibit clear peaks, with a negative peak around $-15^\circ \ \mathrm{to} \ -30^\circ$ for $-y$ motion and a corresponding positive peak for $+y$ motion, indicating a preferred body angle for lateral migration. Rather than being sharply localized, these peaks span a finite range, as turbulence and local flow variability cause the fish's orientation to fluctuate around a preferred angle. 

Orientations opposing the direction of motion occur in only about $13\%$ of cases and are most commonly associated with near-zero orientation or lateral velocity, consistent with station holding rather than active migration. The remaining instances correspond to brief transients as fish adjust their orientation to initiate or halt movement, while cases involving sustained motion against the flow represent only a small fraction. Additional discussion of these minority behaviors is provided in the supplementary material.



\begin{figure*}
\centering
\includegraphics[width=\textwidth]{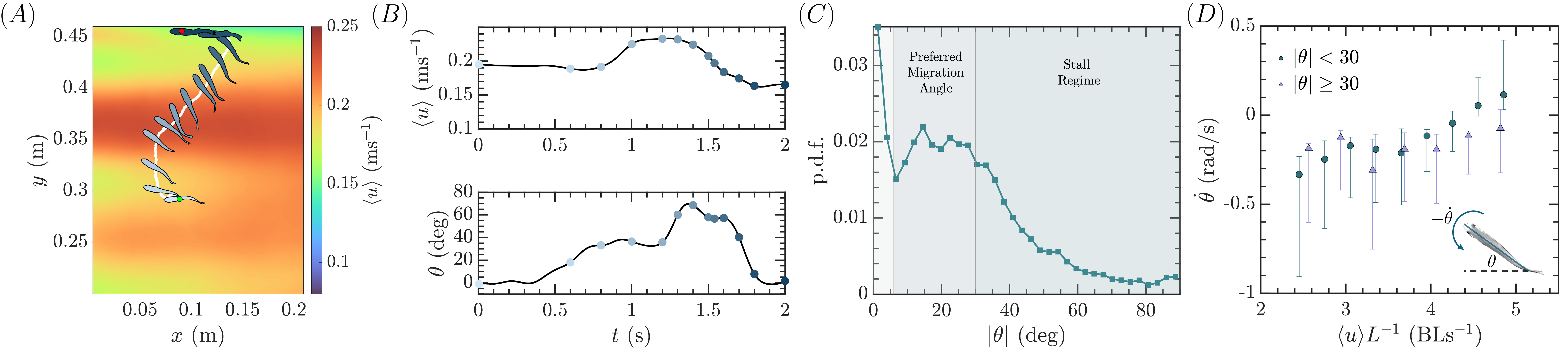}
\caption{\textbf{Yaw dynamics and dependence on flow velocity during lateral migration}. (A) Example trajectory of fish moving away from a high-velocity region to a lower-velocity region before resuming upstream swimming 5 cm away from the wall boundary (\legendcircleG{1.2}{0.1} represents the starting point and \legendcircleR{1.2}{0.1} the end point.) (B) Mean flow velocity along the trajectory and orientation angle as a function of time; markers indicate the time points shown in (A). (C) P.d.f. of the orientation angle for trajectory segments with nearly fixed orientation over more than one tail-beat cycle. The primary peak near $0^\circ$ corresponds to station holding, while secondary peaks of $15^\circ$ and $25^\circ$ correspond to lateral migration. (D) Mean angular velocity versus mean flow velocity normalized by fish body length for two yaw angle ranges, showing flow-dependent yaw dynamics. Symbols in (B) and (C) denote the event-level mean; error bars show 95\% confidence intervals estimated using a hierarchical bootstrap over tracks and events. Track and event counts for each bin is reported in Table S2.}
\label{fig:Fig3}
\end{figure*}

Next, we evaluate whether the observed behavior follows the expected lift force scaling. When the fish yaws its body, it generates a lift force $L$ in the lateral direction, which scales as $C_Lu_{rel}^2 \sim C_L(U_f - u)^2$, where $C_L$ is the coefficient of lift. $u_{rel}$ is the relative velocity, $U_f$ is the absolute fish velocity, and $u$ is the local mean flow velocity at the location of the fish, all in the streamwise direction. Under a quasi-steady approximation, the resulting lateral motion is opposed by a proportional drag-like resistance $D$, which scales as $C_Dv_{rel}^2 \sim C_D(V_f - v)^2$, where $C_D$ is the drag coefficient. $v$ is the local mean flow velocity at the fish's location, and $V_f$ is the absolute fish velocity, both in the spanwise direction.

The morphology of many fishes resembles that of streamlined bodies \cite{Lighthill1960, Lighthill1971, Webb1975} and has often been compared to airfoil-like geometries exhibiting similar hydrodynamic characteristics \cite{Videler1993, Triantafyllou2000, Lucas2020}. Motivated by this analogy, we adopt a simple scaling framework to interpret the observed lateral migration dynamics. Although freely swimming fish exhibit unsteady three-dimensional body deformations, this simplified model provides a first-order description of how yaw angle may influence lateral motion. In this framework, for small angles of attack, we assume $C_L$ scales approximately linearly with $\theta$, while $C_D$ remains nearly constant. Under these assumptions, a scaling balance between lift-like forcing and drag-like resistance then yields:

\begin{equation}
    D \sim L \rightarrow  v_{rel}^2 \sim \theta u_{rel}^2.
    \label{Eq1}
\end{equation}

In Fig. \ref{fig:Fig2}(B), the lateral drag force is plotted as a function of the lift force, averaged over events where fish are not actively propelled upstream. Error bars denote the 95\% confidence interval of the mean. The observed trend of the data points (symbols) is broadly consistent with the expected linear scaling (dashed line) supporting the proposed lift-based lateral transport. However, the data points deviate from the dashed line at the high-lift extremes. These regions occur at large yaw angles and may reflect the onset of stall, where flow separation along the body increases drag and reduces lift. The points where the curve diverges from the expected scaling occur around $\pm 0.056$. The average value of the angles contained in these bins is $\pm 38^\circ$ with a standard deviation of $8^\circ$. Three-dimensional direct numerical simulations of the flow around a stationary fish predict a maximum-lift angle of attack of approximately $\pm45^\circ$ \cite{Zhou2025b}, which is close to the value observed here. 

If a fish were to maintain exactly the predicted stall angle, even a small perturbation could produce a sharp increase in drag and a substantial loss of lift. Under such conditions, the fish would be advected primarily downstream with minimal lateral migration, thereby limiting escape from unfavorable flow regions. Although the linear dependence of the lift coefficient holds up to approximately $\pm 38^\circ$, the preferred migration angle of the fish, as shown in Fig. \ref{fig:Fig2}(A), is roughly $\pm 25^\circ$, which lies about one standard deviation below the mean stall angle. This angle preference is consistent with the possibility that fish maintain orientations that promote lateral transport while avoiding orientations that may initiate the onset of stall.


Beyond orientation, fish may further regulate the  lateral lift force by modulating body shape, as indicated in Fig. \ref{fig:Fig2}(C). When expressed as the relative maximum camber, $p$ (the maximum camber normalized by chord length, or fish body length), negative camber increases the lift force at negative orientation angles but decreases it at positive angles, whereas positive camber produces the opposite effect. These results suggest that, beyond body orientation alone, fish may also modulate the lateral force through body curvature, analogous to adjusting the camber of an airfoil.

The results in Fig. \ref{fig:Fig2}(B) show that orientating near the inferred stall angle is associated with larger  lift forces that help facilitate lateral movement. Beyond this hydrodynamic benefit, we next consider whether maintaining such an orientation might also influence access to flow-related sensory cues. 

A quantity that may contribute to sensing local flow-speed variation is the yaw-induced bilateral pressure difference. Ristroph et al. \cite{Ristroph2015} showed that the canal neuromast density is highest where the bilateral pressure differences are the greatest, suggesting that this quantity may be a relevant sensory cue for the lateral line system. To explore how yaw might influence sensitivity to local flow‑speed changes, we approximate the bilateral pressure difference $\Delta P$ in the pre‑stall regime as follows, where $\rho$ is the fluid density and $K$ is a proportionality constant

\begin{equation}
\Delta P(u,\theta)= P_\mathrm{{right}}-P_{\mathrm{left}}\approx \frac{1}{2}\rho u^2 K\theta ,
\end{equation}

so that the sensitivity to local flow-speed changes is

\begin{equation}
G_u(\theta)=\frac{\partial \Delta P}{\partial u}\approx \rho u K\theta .
\end{equation}

The magnitude of the sensing gain increases linearly with yaw angle and is largest at the upper bound of the pre-stall regime,

\begin{equation}
\begin{gathered}
|G_u(\theta)| \approx \rho u K |\theta|, \\
|G_u| \text{ is maximal as } |\theta| \to \theta_{\mathrm{stall}}^{-}.
\end{gathered}
\end{equation}

This linear dependence on $\theta$ suggests that larger yaw angles amplify the bilateral pressure signal generated by flow-speed variations, potentially enabling fish to detect smaller local flow perturbations. This implies that maintaining a large pre-stall yaw angle may not only enhance lateral migration but also improve hydrodynamic sensing.

Next, we assess whether the observed behavior is consistent with a pressure-based flow-sensing mechanism. To do so, we first tested for evidence of sustained orientation control. Specifically, we extracted trajectory segments in which the change in orientation angle between successive time steps was less than $0.25^\circ$, retaining only continuous segments longer than one tail-beat period ($\sim 0.125\ \mathrm{s}$).

\begin{figure*}[t!]%
\centering
\includegraphics[width=0.90\textwidth]{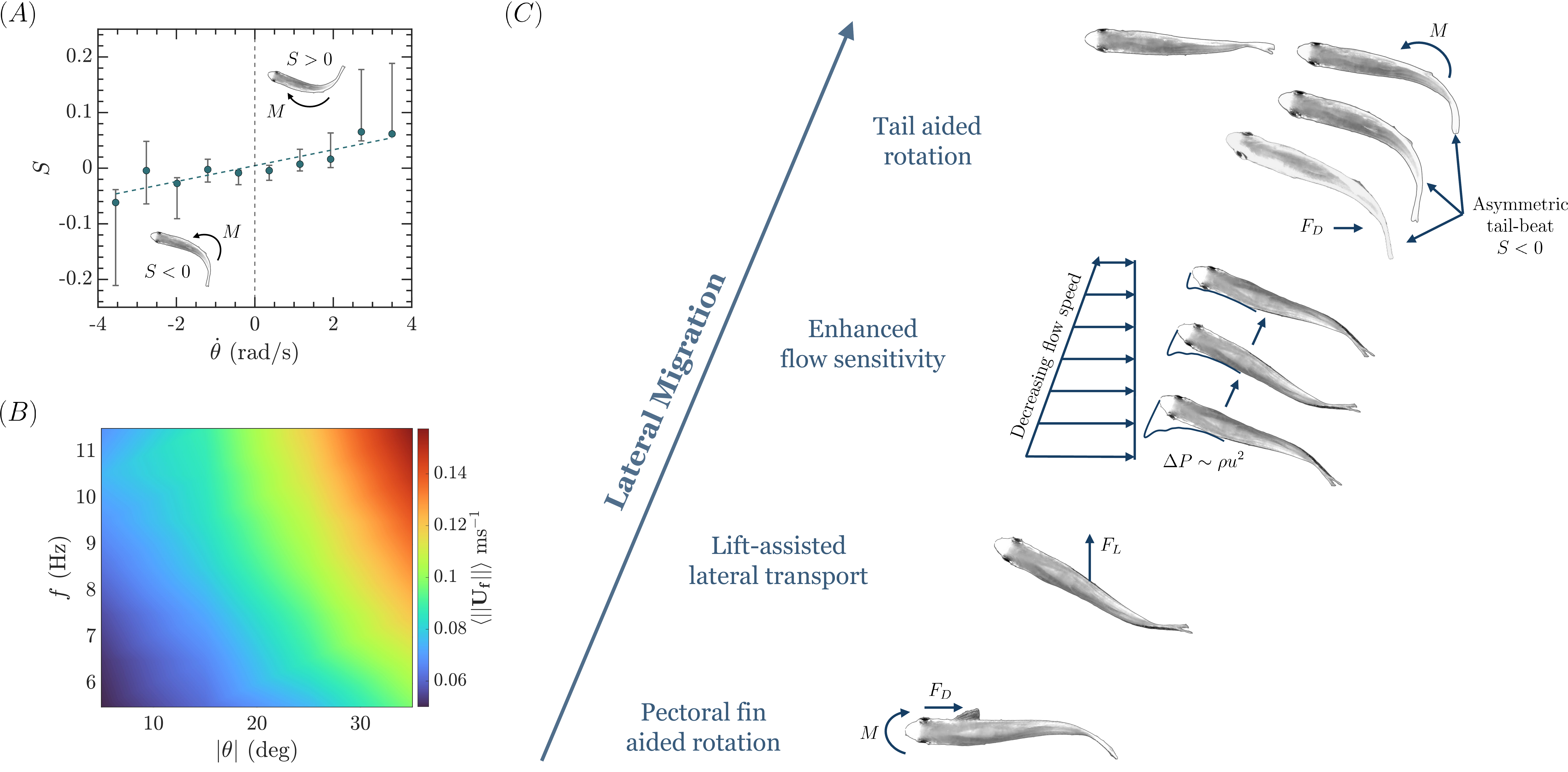}
\caption{\textbf{Body reorientation, lateral motion, and kinematic trends associated with migration through heterogeneous flow}. (A) Stroke asymmetry index $S$ as a function of the angular velocity $\dot{\theta}$, indicating that asymmetric tail beats are associated with body rotation. (B) Average absolute fish velocity versus tail-beat frequency and orientation angle, indicating that higher velocities can be achieved at the same tail-beat frequency by increasing body orientation. (C) Schematic summary of the proposed navigation framework: asymmetric pectoral fin use may help rotate the body and increase orientation, enabling lift-assisted lateral motion and enhanced flow sensitivity; in lower-velocity regions, asymmetric tail beats may help realign the body before upstream swimming resumes. Symbols in (A) denote the event-level mean; error bars show 95\% confidence intervals estimated using a hierarchical bootstrap over tracks and events. Track and event counts for each bin is reported in Table S2.}
\label{fig:Fig4}
\end{figure*}

Fig. \ref{fig:Fig3}(C) shows the p.d.f. of the resulting orientation angles, whose peaks identify the absolute value of the angles most consistently maintained by the fish. The primary peak occurs around $0^\circ$ indicating the fish's preference to station-hold predominately in regions outside of the jets. Secondary peaks occur around $15^\circ$ and $25^\circ$, which agree well with the peaks observed in the p.d.f.s in Fig. \ref{fig:Fig2}(A). Observing that this same range of angles is consistently maintained along trajectories suggests that fish regulate their orientation over extended periods rather than adopting these orientations only transiently. This behavior is consistent with fish using bilateral pressure differences to sense local flow variations to seek favorable hydrodynamic locations. In contrast, there is a steep decline in angles greater than $35^\circ$ being maintained by fish. Based on the results in Fig. \ref{fig:Fig2}(B), at these larger angles, it is expected that lateral lift would decrease, and the flow would become separated and unsteady, potentially leading to messy pressure cues. This may help explain why fish avoid these larger yaw angles.

Having established that fish maintain a preferred range of yaw angles during migration, we next examine whether yaw dynamics are modulated by local flow conditions. If such modulations were present, then individual trajectories should show fish increasing their yaw angle in high-velocity, high-turbulence regions, maintaining it under intermediate conditions, and then reducing it as they enter low-velocity, low-turbulence areas. One such example is shown in Fig. \ref{fig:Fig3}(A)–(B). Starting in a high-velocity region, the fish increases its body angle to help initiate lateral migration. As it moves sideways, it largely maintains this orientation while making small adjustments in response to local velocity changes. After reaching a lower-velocity region, the fish reduces its yaw angle, realigns more closely with the incoming flow, and resumes upstream swimming.

We next asked whether similar structure is observed at the population level. To test this, we identify trajectories in which fish experience a substantial local velocity difference and compute the yaw angle and yaw velocity for instances of active repositioning (orientation greater than $6^\circ$ while moving downstream). Fig. \ref{fig:Fig3}(D) shows that the mean yaw velocity depends on both normalized flow speed and yaw angle. The most positive yaw velocities occur at high flow speeds, indicating a tendency for fish to increase yaw in high-velocity regions to generate lift and migrate away. As flow speed decreases below approximately 3.5 $\mathrm{BLs^{-1}}$, the mean yaw velocity becomes increasingly negative, reflecting a tendency to reduce yaw and realign with the flow in lower-velocity regions. Moreover, as the yaw angle approaches the predicted stall regime, the mean yaw velocity becomes negative across all flow speeds, suggesting that fish actively reduce yaw to remain within the pre-stall range where lift generation and flow sensitivity are expected to be maintained.

Having characterized how yaw varies during migration, we next examine the kinematic mechanisms used to generate these orientation changes. Unlike the rapid, high-acceleration C-start turns used in predator-prey interactions \cite{Dabiri2020,Paniccia2022, Tong2025}, the maneuvers observed here are slower and may be less mechanically demanding. Rather than relying on impulsive tail acceleration, turning appears to arise primarily from asymmetric distribution of drag forces on the body. One possible mechanism is asymmetric deployment of a pectoral fin, which can increase drag on one side and produce a turning moment in that direction \cite{Drucker2001}. A similar effect may also arise from asymmetric body undulations: if tail-beat amplitude is larger on one side, the drag imbalance can also rotate the body \cite{Hoover2020}. 

The imaging system was designed to maximize the field of view and capture long-duration trajectories throughout the flow field. This choice inevitably reduced the spatial resolution available for resolving fine-scale body features, preventing reliable quantification of pectoral-fin kinematics. Although asymmetric pectoral-fin deployment was occasionally observed preceding rotational events, we focus our quantitative analysis on tail-beat kinematics, which can be robustly extracted across all trajectories. Moreover, the caudal fin is expected to play the dominant role in generating maneuvering forces because of its larger size and greater torque-producing capacity. To quantify tail-beat asymmetry, we define a stroke asymmetry index $S$, computed for each tail beat $k$ and averaged across all fish trajectories and swimming conditions:

\begin{equation}
    S = \left\langle \frac{|A_R|-|A_L|}{|A_R| + |A_L|} \right\rangle_k,
    \label{Eq2}
\end{equation}

where $A_R$ and $A_L$ denote the rightward and leftward tail excursions, which is measured using the camber at the tail location. 

For each stroke, the yaw velocity is computed by forward differencing the orientation angle at the beginning and end of the beat period. The data were binned according to yaw velocity, and the stroke asymmetry index was averaged within each bin. Fig. \ref{fig:Fig4}(A) shows that negative yaw velocities are associated with leftward tail excursions ($S < 0$), which generate a counterclockwise moment that reduces the orientation angle. Likewise, positive yaw velocities are associated with rightward tail excursions ($S > 0$). Furthermore, yaw velocities near zero are associated with symmetric tail beats ($S \approx 0$). 

These turning mechanisms can be integrated into a unified navigation strategy, summarized in Fig. \ref{fig:Fig4}(C). In this representative trajectory, the fish first deploys a pectoral fin asymmetrically, consistent with generating a drag imbalance to help rotate the body and increase the orientation angle toward the inferred stall point. This increased orientation is associated with lift-like lateral forcing, which helps to initiate lateral migration. During migration, fish maintain this orientation, potentially allowing the lateral line to detect changes in the pressure difference across the body that may reflect variations in local flow velocity. Upon reaching a low-velocity, low-turbulence region, the fish then uses asymmetric tail beats to generate the differential drag needed to rotate the body and realign with the flow. Once settled, the fish resumes upstream swimming under more favorable flow conditions.

Lastly, we examine whether lateral migration can enhance fish locomotion without requiring a substantial increase in tail-beat activity, which serves as an indirect indicator of locomotor effort \cite{Sanchez2023}. To do so, we analyze the relationship between swimming speed, tail-beat frequency, and body orientation. Fig. \ref{fig:Fig4}(B) shows the average absolute fish velocity ($||\mathbf{U_f}|| = (U_f^2 +V_f^2)^{1/2}$) as a function of tail-beat frequency and orientation angle, averaged over each tail-beat cycle across all trajectories and swimming conditions. At a fixed tail-beat frequency, fish achieve larger absolute velocities at larger orientation angles.

This result suggests that fish can increase their migration speed through body reorientation rather than by substantially increasing tail-beat frequency. Together with the potential for improved detection of favorable flow at these larger body orientations, this may help to explain why lateral migration emerges as a preferred navigation strategy in this hydrodynamic maze.

\section*{Material and Methods}

\subsection*{Animals and Housing}

Giant danio (\emph{Devario aequipinnatus}), with a nominal body length of $\sim 5 $ cm and height $\sim 1$ cm, were procured from a local pet store (PetSmart). The fish were housed in multiple 20 gal tanks. All tanks are equipped with temperature regulation (24-26 \textdegree C), aeration, and filtration systems. Water changes were performed weekly, and fish were fed commercial flake food daily. The animal holding and experimental procedures were approved by the Johns Hopkins Animal Care and Use Committee (IACUC) under protocol number FI24E79.

\subsection*{Flow Characterization}

Experiments were performed in an open water channel that provides a swimming area of $30 \times 50$ cm and depth of $20$ cm. The location of the fish in the streamwise direction was restricted by a pair of low-solidity screens ($14\%$) constructed from $250\ \mathrm{\mu m}$ diameter wire. The hydrodynamic labyrinth was created using a jet array system, which provides a repeating pattern of unfavorable flow conditions through which fish must navigate. Three different flow fields were utilized during this experiment by systematically increasing the jet velocity, which increases the local turbulence intensity and mean velocity in the location of the jets. To characterize the flow in the channel, particle image velocimetry (PIV) was carried out on the vertical midplane without fish. Further details on the experimental facility and flow fields can be found in \cite{Calicchia2026} and the supplemental materials.

\subsection*{Experimental Procedure}

During this experiment, 20 fish were utilized to provide the multiple samples necessary to account for individual variations in behavior. The fish were transferred with a hand net from their housing tanks to the section of the channel that is approximately $20$ cm from the jet array system. The fish were left idle without any flow for a few hours to allow them to acclimate to the new tank and lighting environment. Infrared LED panels were utilized to illuminate the fish on the image plane. During the experiment fish swam continuously across the three swimming conditions, as the jet speed was systematically increased. Fish swam for roughly ten minutes in each flow condition. In the first five minutes, they were allowed to adjust to the flow and in the last five minutes their dynamics were recorded. At the conclusion of the test, fish were allowed to relax for roughly 30 minutes in the channel without flow before being returned to their tanks.   

\subsection*{Visualization}

The fish trajectories were recovered from a ventral-view visualization obtained from a high-resolution camera operating at a frame rate of 250 fps. An in-house MATLAB code was used to track the fish and to extract the orientation angle, midline kinematics, and boundary points. This provides the full spatiotemporal evolution of each fish. A two-dimensional nearest neighbor tracking method was utilized to track the centroids of each fish. Further details of the tracking and pose estimation of the fish is provided in the supplementary material.  

\subsection*{Data Statistics}

The tail-beat frequency, absolute velocities, and orientation angles were extracted for each fish. All points in Fig. \ref{fig:Fig2}(B)-(C), \ref{fig:Fig3}(D), and \ref{fig:Fig4}(A) are event-level means with 95\% confidence intervals obtained from hierarchical bootstrapping over tracks and events. Because individual fish identity could not be maintained continuously across recordings and flow condition changes, tracks were used as the highest available sampling unit. The color map in Fig. \ref{fig:Fig4}(B) show event-level means pooled across all swimming conditions and fish. Additional details on the statistical analysis, as well as track and event counts for each data point, are provided in the supplementary material. The distributions shown in Figs. \ref{fig:Fig2}(A) and \ref{fig:Fig3}(C) are approximate p.d.f.s estimated using normalized histograms, whereas Fig. \ref{fig:Fig1}(E) was estimated via kernel density estimation. Additional details regarding the kernel, sampling choices, and the numbers of events and tracks contributing to each p.d.f. are provided in the Supplementary Material.

\section*{Acknowledgments}

We would like to thank Miguel X. D\'iaz-L\'opez, Matt Gorman, and Xuan Ruan for their guidance in setting up the laser system, Clara Mart\'inez for helping to reorganize the lab in preparation for these experiments, and the Lauder Laboratory for providing the zoomed‑in lateral migration video, which clearly shows all components of this navigation strategy. This work was supported by funds from the Office of Naval Research (N00014-21-1-2661 and N00014-24-1-2152).

\bibliographystyle{unsrtnat}
\bibliography{ref}

\end{document}